\documentclass[useAMS,usenatbib]{mn2e}
\usepackage{graphicx}
\usepackage{rotating,times,pictex,graphicx,latexsym}
\usepackage{color}
\usepackage{longtable,amsmath}
\usepackage{lscape}
\usepackage{lipsum,natbib}

\newcommand{\dg}{$^{\circ}$}

\newcommand{\htwo}{H$_2$}

\title{YSO jets in the Galactic Plane from UWISH2:\\
 I - MHO catalogue for Serpens and Aquila} 

\author[Ioannidis \& Froebrich]{G.~Ioannidis$^{1}$\thanks{E-mail:
gi8@kent.ac.uk}, D.~Froebrich$^{1}$\thanks{E-mail: df@star.kent.ac.uk}\\ $^1$
Centre for Astrophysics and Planetary Science, University of Kent, Canterbury,
CT2 7NH, UK } 

\begin{document}

\date{Received sooner; accepted later}
\pagerange{\pageref{firstpage}--\pageref{lastpage}} \pubyear{2011}
\maketitle

\label{firstpage}

\begin{abstract}

Jets and outflows from Young Stellar Objects (YSOs) are important signposts of
currently ongoing star formation. In order to study these objects we are
conducting an unbiased survey along the Galactic Plane in the 1-0\,S(1) emission
line of molecular hydrogen at 2.122\,$\mu$m using the UK Infrared Telescope. In
this paper we are focusing on a 33 square degree sized region in Serpens and
Aquila (18\dg \textless l \textless 30\dg; -1.5\dg \textless b \textless
+1.5\dg). 

We trace 131 jets and outflows from YSOs, which results in a 15 fold increase in
the total number of known Molecular Hydrogen Outflows. Compared to this, the
 total integrated 1-0\,S(1) flux of all objects just about doubles, since
the known objects occupy the bright end of the flux distribution. Our
completeness limit is 3\,$\cdot$\,10$^{-18}$\,W\,m$^{-2}$ with 70\,\% of the
objects having fluxes of less than 10$^{-17}$\,W\,m$^{-2}$. 

Generally, the flows are associated with Giant Molecular Cloud complexes and
have a scale height of 25\,--\,30\,pc with respect to the Galactic Plane. We are
able to assign potential source candidates to about half the objects. Typically,
the flows are clustered in groups of 3\,--\,5 objects, within a radius of 5\,pc.
These groups are separated on average by about half a degree, and 2/3rd of the
entire survey area is devoid of outflows. We find a large range of apparent
outflow lengths from 4\arcsec\ to 130\arcsec. If we assume a distance of 3\,kpc,
only 10\,\% of all outflows are of parsec scale. There is a
2.6\,$\sigma$ over abundance of flow position angles roughly perpendicular
to the Galactic Plane.

\end{abstract}


\section{Introduction}

The interstellar medium (ISM) in galaxies is radically influenced by star
formation. Giant Molecular Clouds (GMCs) are heated and excited by outflows from
protostars and radiation from high-mass young stellar objects (YSO). Changes in
chemistry and probably the turbulent motion in GMCs are a result of star
formation, especially massive star formation. Therefore, it is of great
importance to understand the formation of stars.

Jets and outflows from YSOs are sign-posts of currently ongoing star formation
(e.g. \cite{Bally1995}; \cite{Eisloffel2000}; \cite{Froebrich2003};
\cite{Davis2009}). Previous studies of star forming regions like the
Orion\,A molecular ridge by \cite{Davis2009} and \cite{Stanke2002}, DR\,21/W\,75
by \cite{Davis2007}, as well as the Taurus-Auriga-Perseus clouds by
\cite{Davis2008}, have shown a large number of outflows from YSOs. However,
there are a number of open questions which are still to be addressed. For
example: Is a large number of jets and outflows a common occurrence in
other star forming regions (low and high mass)? Is star formation triggered in
infrequent bursts or is it an ongoing, multiple epoch process in each GMC? The
presence of jets from YSOs is an indication of a young population and active
accretion while the sparsity of them in regions with a sizeable population of
reddened sources shows a more evolved region with a larger population of
pre-main-sequence stars. The dynamical age of a protostellar outflow is 10 to
100-times less than the turbulent lifetime of a GMC. Therefore, the presence
of a large number of outflows will be an indication of currently ongoing or
multiple epochs of star formation.

Outflows from YSOs are also a direct tracer of mass accretion and ejection and
can be used to estimate star formation efficiency from region to region. 
This is particularly the case in high mass star forming regions where the
efficiency is grossly affected by existing massive young stars, which influence
the environment via their hugely energetic winds and intense UV fluxes.
Furthermore, in massive star forming regions, where young stars form in clusters
and massive stars influence their lower-mass neighbours, photo-evaporation and
ablation of protostellar disks can suppress accretion. As a result the mechanism
that drives the jets and outflows is switched off. To what degree do these
interactions affect accretion in YSOs?

The position of individual protostars can be detected from jets and outflows
while their evolutionary stage (e.g. Class\,0, Class\,I) is closely related to
the brightness in \htwo\ emission \citep{CarattioGaratti2006a}. The mass
infall/ejection history can be determined from jets since there is a correlation
between the jet parameters, mass infall rates and accretion luminosities
\citep{Beck2007, Antoniucci2008}.

Studies by \cite{Eisloffel1994} and \cite{Banerjee2006} suggest that outflows
are aligned parallel with the local magnetic field and perpendicular to the chains of
cores. However, existing observations give mixed results without a definite
answer \citep{Anathpindika2008, Davis2009}. Is there a correlation between the
mass of the driving source and the orientation of the flow with respect to the
cloud filament (i.e. are massive outflows more likely to be orthogonal to the
filaments than flows from low mass stars)? Furthermore: What fraction of
outflows and jets are collimated, and what fraction are parsec-scale in length?
Is there a correlation between the median age of the embedded population and the
mean flow length? Are outflows sufficient in numbers and energetic enough to
account for the turbulent motions in GMCs?

In order to answer the above outlined questions we need to have a representative
sample of jets and outflows from young stars which is free from selection
effects. This will allow us to perform a statistically meaningful investigation
of the dynamical processes associated with (massive) star formation. Therefore
we are conducting an unbiased search for jets and outflows in the Galactic Plane
using the UKIRT Wide Field Infrared Survey for H$_2$ (UWISH2 --
\cite{Froebrich2011}, hereafter F11). The data is taken in the \htwo\
1-0\,S(1) emission line at 2.122\,$\mu$m, and thus highlights regions of shock
excited molecular gas (T $\approx$ 2000\,K, n$_{\rm H_2} >$\,10$^3$\,cm$^{-3}$),
as well as fluorecently excited material. Hence, it can be used to trace
outflows and jets from embedded young stars.

In this project we focus our attention on the Serpens/Aquila region in the
Galactic Plane, covered by UWISH2. In particular we investigate the area 18\dg
\textless l \textless 30\dg; -1.5\dg \textless b \textless +1.5\dg, which
approximately covers 33 square degrees (see Fig.\,\ref{positions} for an
overview of the region). It is the first continuous area of this size completed
and covers about 20\,\% of the total UWISH2 area. Hence, it will allow us to
obtain first but statistically significant results about the population of jets
and outflows from young stars in the Galactic Plane.

In this paper we present our analysis of the above mentioned region. We focus on
the detection of the jets and outflows, as well as their potential sources. We
further analyse the spatial distribution of the discovered objects, their
apparent lengths, position angles and fluxes. To convert apparent
measurements (such as size) into physically meaningful measurements we need the
distance. We assume that all outflows are at a distance of 3\,kpc throughout the
paper, since this turns out to be the most commonly measured distance for outflows in our
sample. In our forthcoming paper (Ioannidis \& Froebrich, in prep., Paper\,II)
we will discuss in detail our distance measurement method of the individual jets
and outflows. We will then determine in detail statistically corrected
luminosity functions and length distributions of the jets and outflows and
investigate their distribution in the Galactic Plane. Finally, in Ioannidis \&
Froebrich (in prep., Paper\,III) we will investigate the driving source
properties as well as the environment (cloud structure, clustered and isolated
star forming regions) the jets and outflows are in.

In Sect.\,\ref{dataandanalysis} we discuss our data and analysis methods. We
then present our results and discussion in Sect.\,\ref{results}, which includes
the positions and distribution of outflows, the lengths and flux distributions
as well as their orientation. We conclude our findings in
Sect.\ref{conclusions}.

\section{Data and Analysis}\label{dataandanalysis}

\subsection{Near infrared UKIRT WFCAM data}\label{data}

We obtained our near infrared narrow band imaging data in the 1-0\,S(1) line  of
H$_2$ using the Wide Field Camera (WFCAM - \cite{Casali2007}) at the United 
Kingdom Infrared Telescope (UKIRT). The camera consists of four Rockwell
Hawaii-II (HgCdTe 2048\,$\times$\,2048) arrays with a pixel scale of 0.4\arcsec.
The 1-0\,S(1) filter is centred at 2.122\,$\mu$m with
$\Delta\lambda$\,=\,0.021\,$\mu$m. Our data is part of the UWISH2 survey (see 
F11 for details) and the H$_2$ images are taken with a per pixel integration
time of 720\,s under very good seeing conditions. The typical full width half 
maximum (fwhm) of the stellar point spread function is 0.7\arcsec, the 
5\,$\sigma$ point source detection limit is about 18\,mag (in broad-band K) and
the surface brightness limit is about 10$^{-19}$\,W\,m$^{-2}$\,arcsec$^{-2}$
when averaged over the typical seeing (F11). Our narrow band  data were taken
between 31st of July 2009 and 9th of September 2010.

Furthermore, we utilised the UK Infrared Deep Sky Survey (UKIDSS) data in the
near infrared $JHK$ bands taken with the same telescope, same instrumental
set-up and tiling as part of the Galactic Plane Survey (GPS, \cite{Lucas2008})
in order to perform the continuum subtraction (H$_2$-$K$) of our narrow band
images and to generate three band colour images ($JHK$, $JH$H$_2$, $JK$H$_2$;
see  Sect.\/\ref{color}). When compared to our narrow band data the NIR broad
band  data has very similar quality (fwhm and depth). Both data sets are taken
typically about 2.5 to 4.0 years apart.

Data reduction and photometry for both NIR data sets are done by the Cambridge 
Astronomical Survey Unit (CASU). Reduced images and photometry tables are 
available via the Wide Field Astronomy Unit (WFAU). The basic data reduction 
procedures applied are described in \cite{Dye2006}. Calibration (photometric 
as well astrometric) is performed using 2MASS (\cite{Skrutskie2006}) and the 
details are described in \cite{Hodgkin2009}.

\begin{figure*}
\begin{center}
\includegraphics[width=17.5cm]{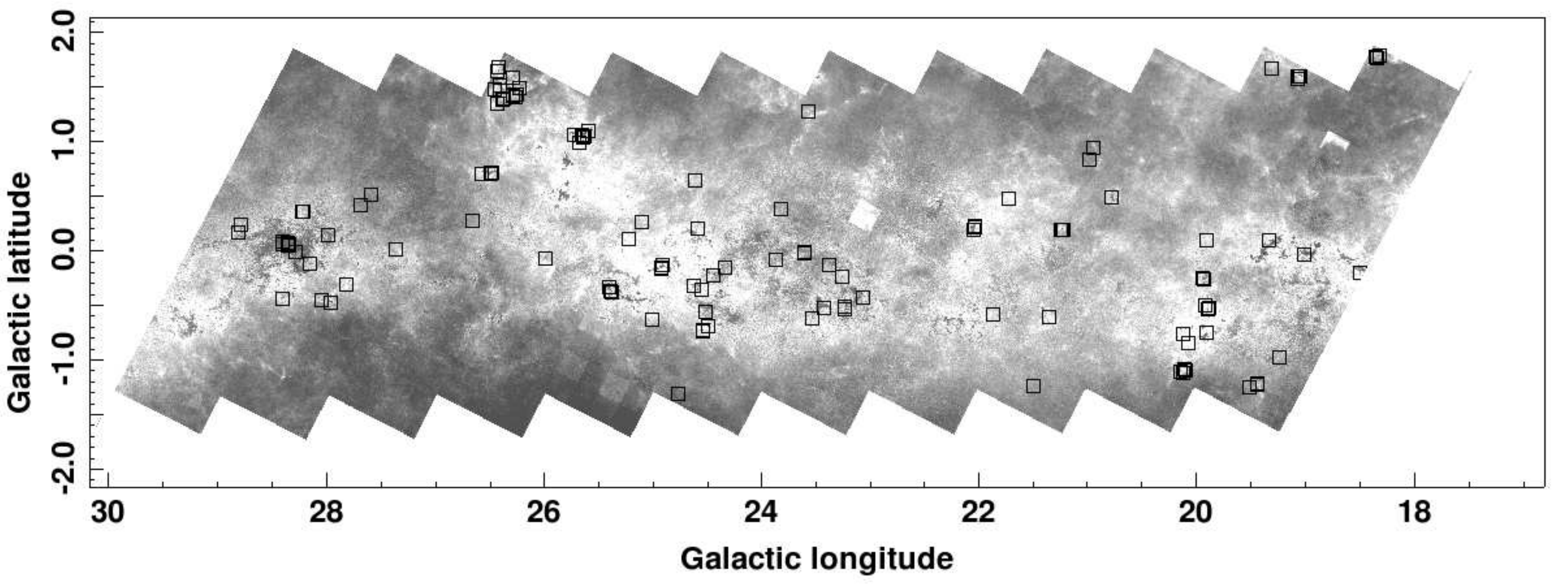}
\end{center}
\caption{\label{positions}Positions of outflows (black squares) over plotted on
a relative extinction map (grey scale image) based on near infrared colour
excess in GPS data. Please see text for details on the background map.}
\end{figure*}

\subsection{Difference Images}\label{difference}

In order to reveal mutually exclusive emission line regions associated with 
H$_2$ jets and outflows (or other H$_2$ emitters), the H$_2$ narrow band images
have been continuum-subtracted using the $K$-band images from the GPS. Note that
we are working with images about 13.3\arcmin\,$\times$\,13.3\arcmin\  
in size throughout the project,
since these are the images delivered by CASU. Every set of images (H$_2$, $K$)
is aligned based on their World Coordinate System (WCS). Photometry is performed
to determine the mean fwhm of each image.

To compensate for the different seeing conditions in both images and to  enhance
the signal to noise ratio in the difference images we Gaussian  smoothed both
images before the continuum subtraction. The image with the  worse seeing
(image $b$) is smoothed using a Gaussian with a fwhm  of $r_s^b$\,=\,0.4\arcsec.
The image with the better seeing (image $a$) is smoothed with
a Gaussian of a larger fwhm  ($r_s^a$), which we calculate as:

\begin{equation*}
r_s^a = \sqrt{ (fwhm^b)^2 - (fwhm^a)^2 + (r_s^b)^2}
\end{equation*}

where $fwhm^a$ is the mean fwhm of image $a$ and $fwhm^b$ is the mean fwhm of 
image $b$. Typically the seeing in both images is very similar and the smoothing 
radii $r_s^a$ and $r_s^b$ are thus not very different.

In order to completely remove continuum sources such as stars in the difference
images, the $K$-band images need to be scaled. Since the scaling factor $sc$
depends on the foreground extinction to the stars, we expect this to vary
significantly across images taken in the Galactic Plane. We thus use a position
dependent scaling factor $sc(x,y)$ ('scaling image') for the continuum
subtraction.

For this purpose we determine the mean scale factor for all stars in small 
sub-regions (10\arcsec $\times$ 10\arcsec). Before applying the scaling image to the K-band it is 
smoothed with a radius of 30\arcsec. Thus, we trace as much structure of the
foreground extinction as possible while ensuring a sufficient signal to noise of
the scale factor.

The final difference images are then created as:

\begin{equation*}
{\rm Difference}_{H_2-K} = H_2 - K * sc
\end{equation*}

Due to the Gaussian smoothing to adapt the stellar fwhm, the resulting 
difference images have a higher signal to noise than the original H$_2$ and 
$K$ images. While most of the non saturated stars are not present the H$_2$ 
emission features are preserved. We note that the difference images are only 
used for the purpose of detection of jets and outflows from protostars (see 
Subsection\,\ref{search}) and not for photometry.

\subsection{NIR Colour images}\label{color}

The search for H$_2$ emission features (as described in Section\,\ref{search}) 
is mainly based on the H$_2$-$K$ difference images. However, the verification 
that the observed feature is indeed a jet or outflow from a young stellar 
object is greatly facilitated by the additional use of near infrared colour 
composite images obtainable from our data sets. These images are also 
extremely useful to identify potential driving sources for the discovered jets 
and outflows.

We thus created full resolution near infrared $JHK$, $JH$H$_2$ and $JK$H$_2$ 
colour composites for every image. Each of the colour combinations enhances 
different aspects of the spectrum and therefore makes it easier to visually 
detect and/or verify objects of interest. More specifically, the $JK$H$_2$ 
images can be used to detect pure H$_2$ emission regions. Since the H$_2$ and 
$K$ images have been taken at two different epochs, very 'red' or very 'green' 
objects can be identified as $K$-band variable stars in the $JK$H$_2$ 
composites. In contrast, the $JHK$ and $JH$H$_2$ images can be used to detect 
objects with $K$-band excess emission, in other words candidate young stellar 
objects.

\subsection{Outflow detection}\label{search}

Our aim to investigate an unbiased sample of jets and outflows from young  stars
in the Galactic Plane cannot just be achieved by performing an unbiased  survey.
We also need to take great care not to introduce any detection bias  when
identifying the objects in our data. Thus, we follow the strategy described
below to find all potential jet and outflow candidates.

All H$_2$-$K$ difference images have been visually inspected for extended 
H$_2$ emission features in full resolution. The 
order of inspection was completely random to avoid the introduction of biases 
due to our search pattern. The original search has been performed by just one 
of us. However, the subsequent cleaning of this input catalogue (see below) 
has been done by two people.

Every detected emission feature that has been identified in the H$_2$-$K$ 
difference images at first has been confirmed in the corresponding H$_2$ image 
to avoid the inclusion of image artefacts. We then checked it against the
full  resolution $JK$H$_2$ colour composite of the region (such as the one shown
in Fig.\,\ref{photometry}). Thus, any object that  could potentially be an
image artefact has been removed from the source list.

Finally, we need to clean the remaining objects from real emission 
contaminants. These include e.g. Planetary Nebulae (PN), Supernova Remnants 
(SNR) and fluorescently excited regions such as edges of molecular clouds, HII 
regions and areas around young embedded clusters with massive stars. Planetary 
Nebulae can usually be identified by their visual appearance. In the case of 
SN shocks and fluorescently excited cloud edges this is more complicated, as 
they can mimic shocks from jets and outflows. We searched the SIMBAD database 
for known SNRs near to our identified objects. Any potential feature detected 
near such known SNRs has been excluded from our list. We inspected regions 
with (obviously) fluorescently excited cloud edges very carefully and excluded 
any object that potentially was not a jet/outflow feature.

The resulting list of H$_2$ emission line objects is thus complete up to the 
survey detection limit, contains (almost) no false positives and is unbiased.

\subsection{Driving sources of outflows}\label{sources}

One vital task in order to understand the properties of the detected jets and 
outflows (e.g. the length distribution) is the identification of their 
potential driving sources. We utilise a number of published catalogues of YSOs 
as well as mid and far infrared source lists for this purpose. These include 
the catalogue of YSOs compiled by \cite{Robitaille2008} from the Spitzer GLIMPSE
survey \citep{Churchwell2009}, detections in the AKARI/IRC mid-infrared
all-sky survey bright source catalogue \citep{Ishihara2010,Yamamura2009}, 
detections in the IRAS Point and Faint Source Catalogue \citep{Moshir1989,
Moshir1991} and detections in the Bolocam Galactic Plane Survey
\citep{Aguirre2011}, which all cover our entire survey field. Additionally we
used $K$-band excess sources identified from $JHK$ detected objects in the
UKIDSS GPS, as well as $K$-band variable sources identified in the H$_2$-$K$
difference images as potential driving source candidates.

To decide on the most likely source for each H$_2$ feature, we over-plot all 
potential source candidates over the H$_2$-$K$ difference images. Based on the 
vicinity of source candidates and H$_2$ emission, the alignment of a number of 
emission knots with a potential source, or the shape of bow-shock like features
we grouped the H$_2$ emission objects into outflows and assigned one MHO number
to each of them, following the procedure outlined in \cite{Davis2010a}. 
Figure\,\ref{photometry} shows MHO\,2204 as an example of a detected jet with a
potential source candidate.

There are cases with several potential source candidates. If it is impossible 
to decide on one specific one we consider all possibilities. In cases where 
non of the above mentioned catalogues allowed us to find a potential source 
for an H$_2$ feature, we additionally searched the SIMBAD database for other 
indicators of YSO outflow sources, such as masers and (sub)-mm sources. 
Finally, if there are several source candidates which apparently are the same 
object (such as an IRAS, Bolocam and Glimpse detection at roughly the same 
coordinates), we use as the source position and identifier the object in the 
survey with the highest spatial resolution. There are a number of cases where 
we clearly detect a small, bipolar and symmetric jet but no object has been 
detected in any survey at the suspected source position. In these cases we 
assume the source to be situated between the H$_2$ features on the flow 
axis, and it is given the identifier 'Noname'. Note, that MHO\,2444 would be 
one of these objects, if it had not had deep JCMT data from 
\cite{DiFrancesco2008}.

\subsection{Photometry}\label{photom}

After the detection of the H$_2$ emission features we have to measure their 
fluxes. This is straight forward, since our H$_2$ images have been flux 
calibrated by CASU.

\begin{figure*}
\includegraphics[width=8.5cm]{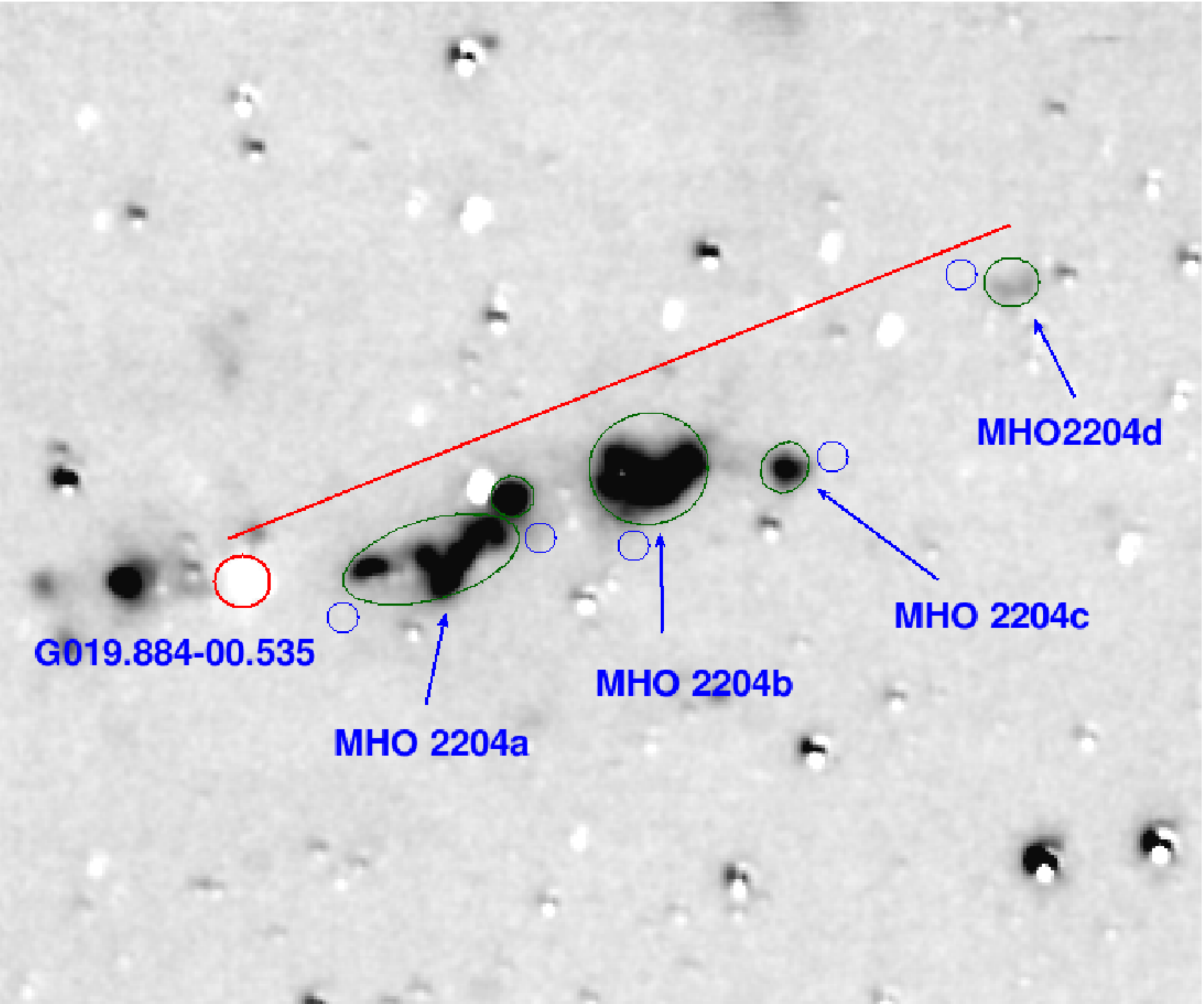}\hfill
\includegraphics[width=8.5cm]{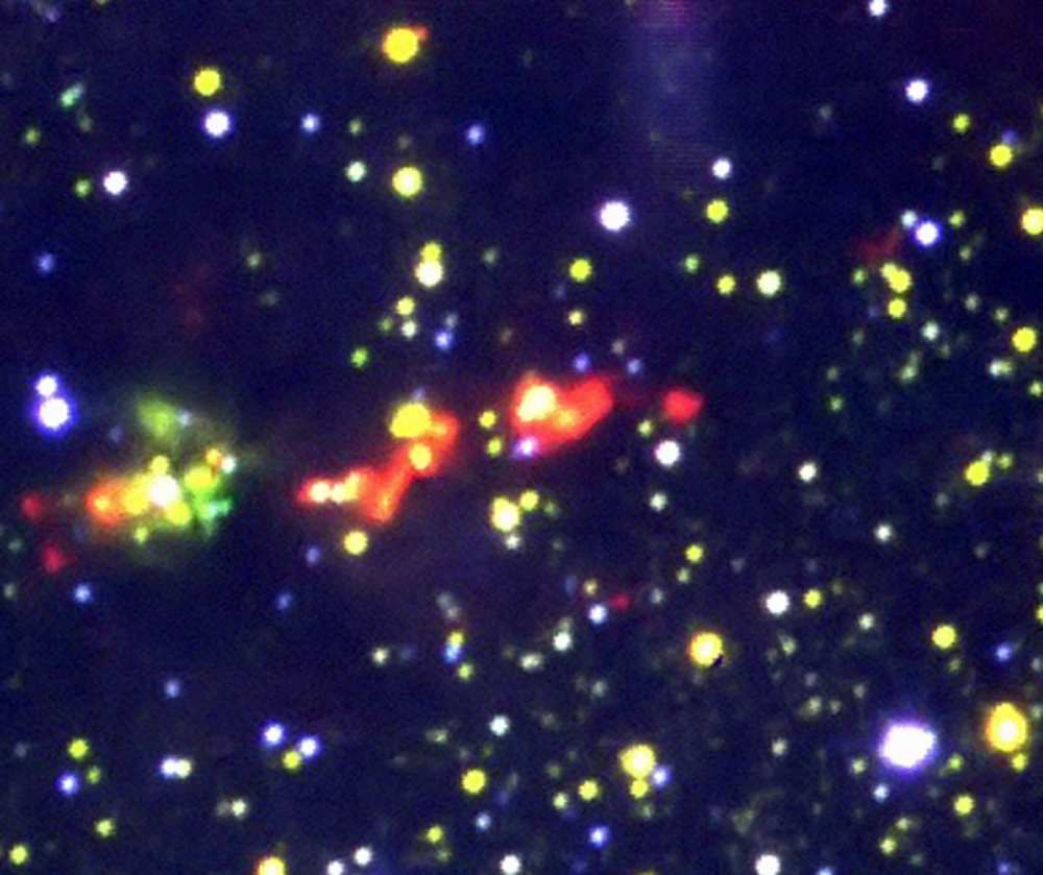}
\caption{\label{photometry} Left: Grey scale representation of the
H$_2$-$K$ difference image of MHO\,2204. The green ellipses indicate the
apertures used to enclose the H$_2$ flux, while the blue apertures represent the
nearby sky. The solid line indicates the direction of the flow. Right: JKH$_2$
color composite of the same region. Both images are 95\arcsec $\times$
66\arcsec\ in size, North is to the top and East to the left. }
\end{figure*}

We define apertures around each emission feature in the H$_2$ images. Great 
care is taken that continuum sources (such as fore or background stars at the 
same line of sight) are not included in the apertures. We also ensure that the 
apertures contain as little area as possible that seems to be free of emission 
in order not to add noise. Simultaneously, we use for each H$_2$ feature a
nearby sky-aperture that is completely emission free and defines the local sky
level (see Fig.\,\ref{photometry} for an example of the apertures used in the
photometry of MHO\,2204). We then measure the number of counts in the H$_2$
feature, corrected for the local background level and repeat that measurement,
using identical aperture positions, in the scaled (using  $sc(x,y)$, see
Sect.\,\ref{difference} above) $K$-band image to correct for the continuum
contained in  the narrow band data.

The background- and continuum-corrected H$_2$ counts for each emission feature 
are then converted into physical flux units ($W/m^2$) using the flux zero 
points provided by CASU. The final flux values for the H$_2$ emission in the 
1-0\,S(1) line (summed up over all knots in each MHO) are shown in 
Table\,\ref{outflow_table} in the Appendix.

Uncertainties in the photometry are based on the variation of the local 
background level in the H$_2$ and $K$-band images and the choice of the exact 
position of the apertures enclosing the H$_2$ emission regions. Typical 
uncertainties of the measured fluxes are discussed in Sect.\,\ref{resflux}.

\section{Results and Discussion}\label{results}

We have detected a total of 131 molecular hydrogen outflows from YSOs in our
survey field of 33 square degrees. Out of these, 121 (92\,\%) objects are new
discoveries. Of the ten already known outflows, two have been recently
identified by Lee et al. (2011 in preparation) using UWISH2 data as well. Thus,
94\,\% of our outflows are newly discovered in our data set. Hence, in the field
investigated here, the UWISH2 data increased the number of known molecular
hydrogen outflows by a factor of 15.

In Table\,\ref{outflow_table} in the Appendix we list the assigned MHO numbers,
positions, fluxes, apparent lengths, position angles, source candidates and
their positions. In Table\,\ref{images_table} we show H$_2$-$K$ images of each
MHO and give a brief description of its morphology and potential sources. We
note that there is a difference between the number of outflows (131) and the
number of MHOs (134). The reason for this is that some of the already assigned
MHO numbers are part of the same outflow. More specifically MHO\,2206, MHO\,2207
and MHO\,2208 are part of the same outflow. Similarly, the bow shock MHO\,2212
is part of the same flow as MHO\,2201. In cases like these we treat all MHOs as
one outflow. All objects in Table\,\ref{outflow_table} that are part of an
outflow containing several MHOs are marked with an asterisk.

\subsection{Spatial distribution and clustering of outflows}

We show the positions of all detected outflows in our field in
Fig.\,\ref{positions}. The grey-scale background map in this figure is a
relative extinction map based on median near infrared colour excess (e.g.
\cite{Rowles2009}) determined from UKIDSS GPS data (\cite{Lucas2008}). Note that
some regions in the map have small A$_V$ off-sets. This is caused by known,
minor photometric calibration issues of the GPS data which will be corrected in
future data releases. Since we are not using the actual A$_V$ values in the
paper, this is a pure 'cosmetic' issue.

As expected, the outflows are mainly located in areas with high extinction.
Their overall distribution hence follows nicely the giant molecular cloud
complexes visible in the A$_V$ map. Only a small number of objects is not
situated within dense clouds. However, there are numerous areas of high
extinction where no outflows have been detected. In Paper\,III we will
investigate in detail if there is an A$_V$ threshold for the detection of
molecular hydrogen flows, and which fraction of high column density clouds
does not show signs of on-going star formation in the form of jets and
outflows.

\begin{figure*}
\includegraphics[width=8.5cm]{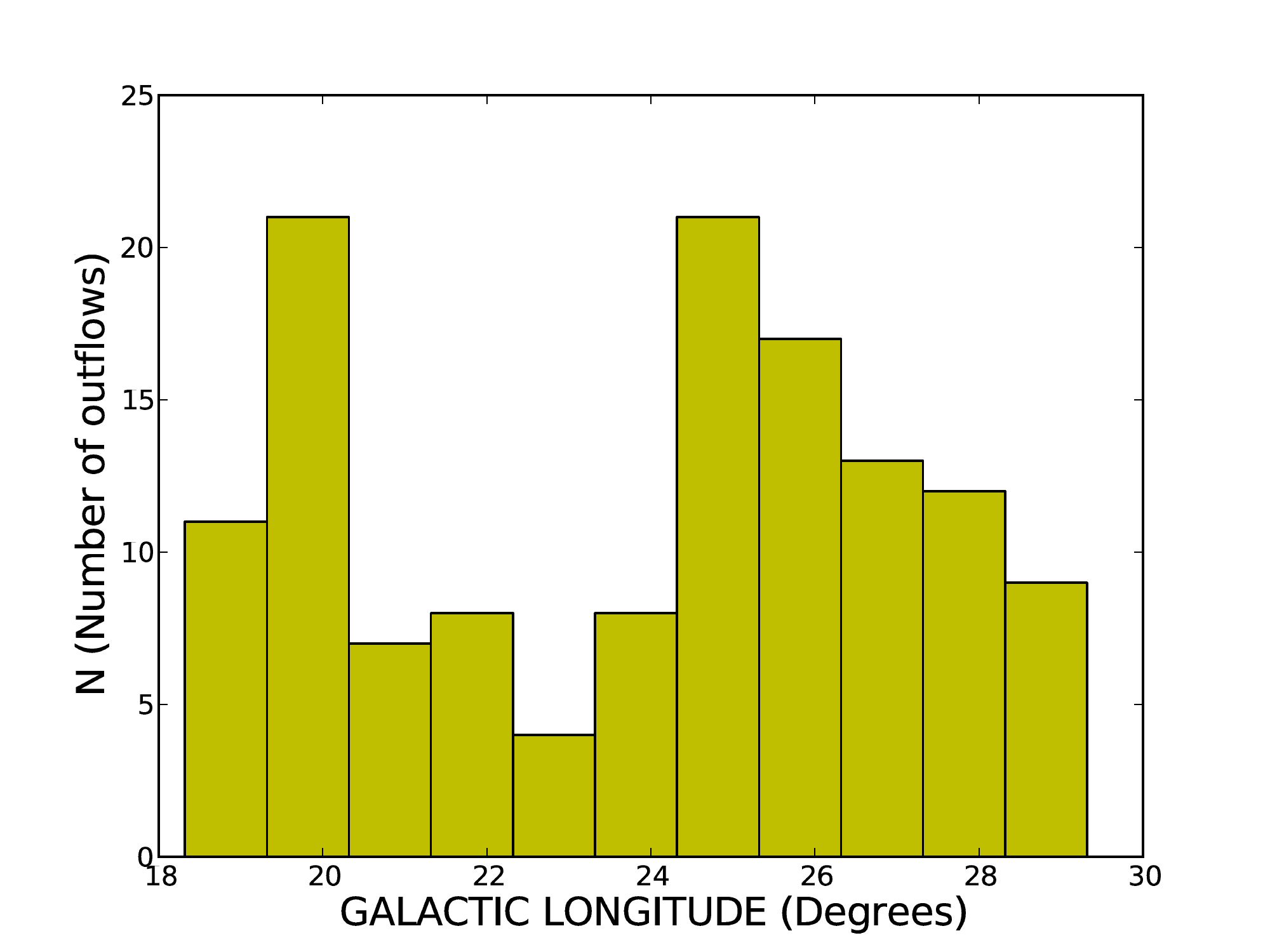} \hfill
\includegraphics[width=8.5cm]{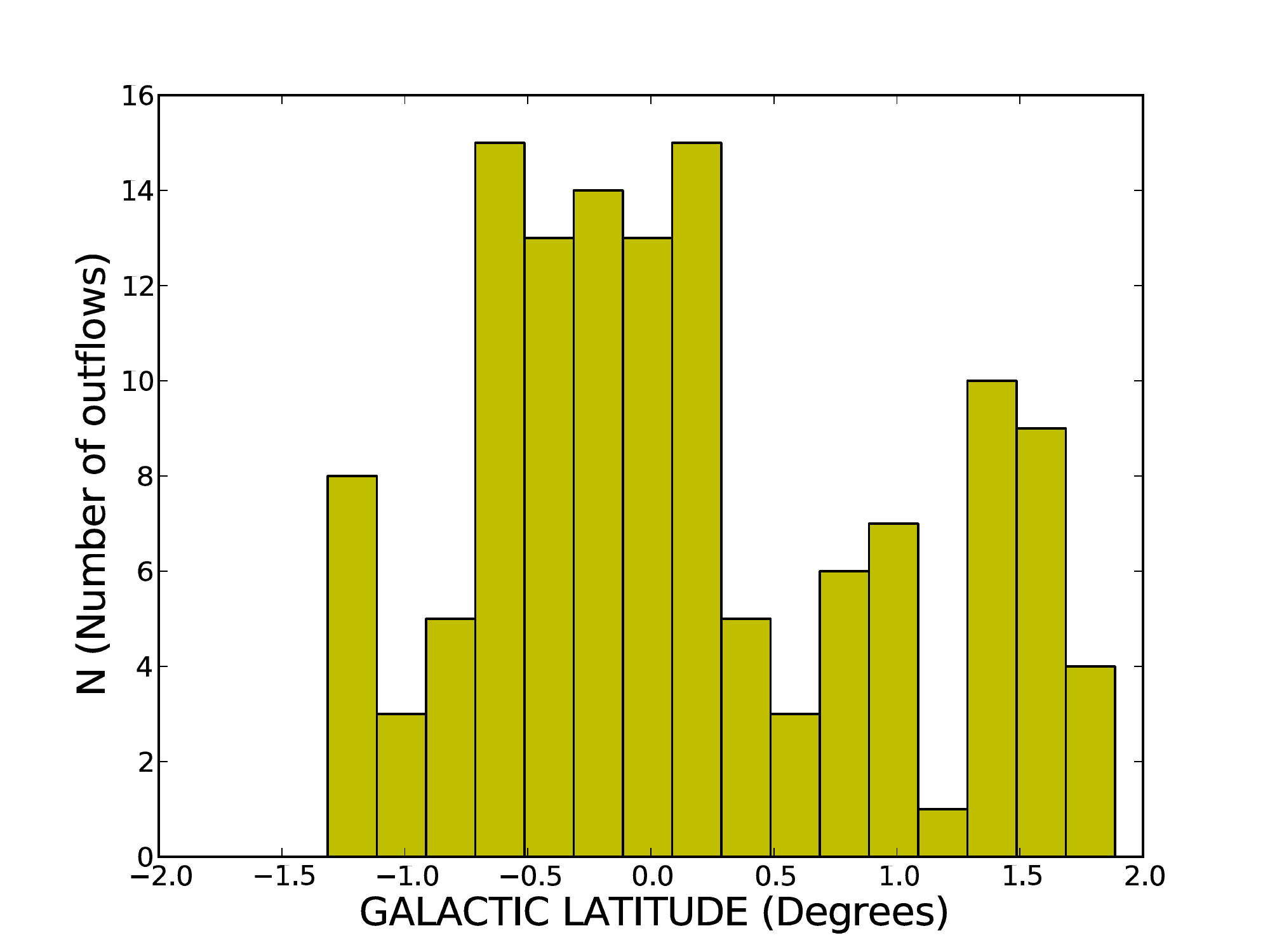}
\caption{\label{lb} Histogram of the positions of the detected outflows along
(left panel) and across (right panel) the Galactic Plane.}
\end{figure*}

Histograms of the outflow positions along and across the Galactic Plane are
shown in Fig.\,\ref{lb}. As one can also see in Fig.\,\ref{positions}, there is
a non homogeneous distribution along the Plane, with peaks at $l$\,=\,20$^\circ$
and $l$\,=\,25$^\circ$ and a minimum around $l$\,=\,23$^\circ$. This indicates
the concentration of outflows to specific areas i.e. the giant molecular cloud
complexes. The latitude distribution shows a Gaussian like distribution with a
width of about one degree. Its centre is shifted to about $b$\,=\,-0.25$^\circ$
hinting that the main cloud complexes in this part of the Galactic Plane are at
negative latitudes. If we assume a distance of 3\,kpc for the outflows, the
width of the distribution corresponds to a scale height of about 25-30\,pc. 
This is very similar to values for young stellar clusters (55\,pc,
\citep{Friel1995}) and OB stars   (30\,--\,50\,pc, \citep{Reed2000, Elias2006}),
but significantly smaller than the about 125\,pc found for the dust  at the
solar galactocentric distance (e.g. \cite{Drimmel2003, Marshall2006}). The
prominent increase of the number of objects at $b$\,=\,1.5$^\circ$ (right panel
of Fig.\,\ref{lb}) is caused by a number of higher latitude clouds in particular
at $l$\,=\,26.5$^\circ$ (see also Fig.\,\ref{positions}).

We have investigated the clustering of the detected outflows by means of nearest
neighbour distance (NND) distributions. Hence, we calculated the distance
to the N$^{th}$ nearest neighbour for all outflows in our sample and determined
a histogram to analyse these distance distributions. The 1$^{st}$, 2$^{nd}$ and
3$^{rd}$ NNDs all show a sharp peak at small separations, indicating that the
flows are typically found in close proximity to each other. This peak almost
disappears in the 4$^{th}$ NND distribution and is certainly gone in the
5$^{th}$ NND distribution, where it is replaced by a wider peak at about
20-30\arcmin\ separation. 

Thus, we find that the outflows typically occur in groups with a few (3-5)
members. The size of these groups is about 6\arcmin\  (a typical value for the
3$^{rd}$ NND), which corresponds to about 5\,pc if we assume all objects are at
a distance of 3\,kpc. This size is clearly larger than the typical embedded
cluster (about 1\,pc or slightly smaller; e.g. \cite{Lada2003}) and smaller than typical nearby GMCs such as
Taurus and Orion. Thus, currently ongoing star formation occurs in those 5\,pc
sized regions within GMCs and is not necessarily confined to embedded clusters.

The groups of outflows are separated by about half a degree on the sky from each
other. Given the number of members in these groups, the average group
separation, the total number of objects and the field size, we can estimate that
about two thirds of the survey area are more or less devoid of molecular
hydrogen emission line objects, a fact supported by the distribution shown in
Fig.\,\ref{positions}.

\subsection{Driving source candidates}

We have been able to assign driving source candidates, as described in
Sect.\,\ref{sources}, to 68 of our 131 molecular hydrogen outflows. This
corresponds to just over half the objects. Overall, 75 source candidates have
been identified, since some outflows have more than one probable driving
source. A complete list of the outflow driving source candidates is presented
in Table\,\ref{outflow_table} in the Appendix.


We find that typically brighter MHOs (according to their integrated
1-0\,S(1) flux, see Sect.\,\ref{resflux} below) are more likely to have a
source candidate assigned to them. In particular the average flux of MHOs with
an assigned source candidate is almost five times higher than for MHOs without
source candidate. For the median fluxes this ratio, however, is lower by a
factor of three. In a future paper we will investigate if this is a pure
selection effect. It could simply be that bright MHOs are close-by and hence
their sources are easier to detect. On the other hand, bright MHOs could be
intrinsically luminous and thus be driven by brighter, easier to detect sources.

\subsection{Apparent outflow lengths}

\begin{figure}
\includegraphics[width=8.5cm]{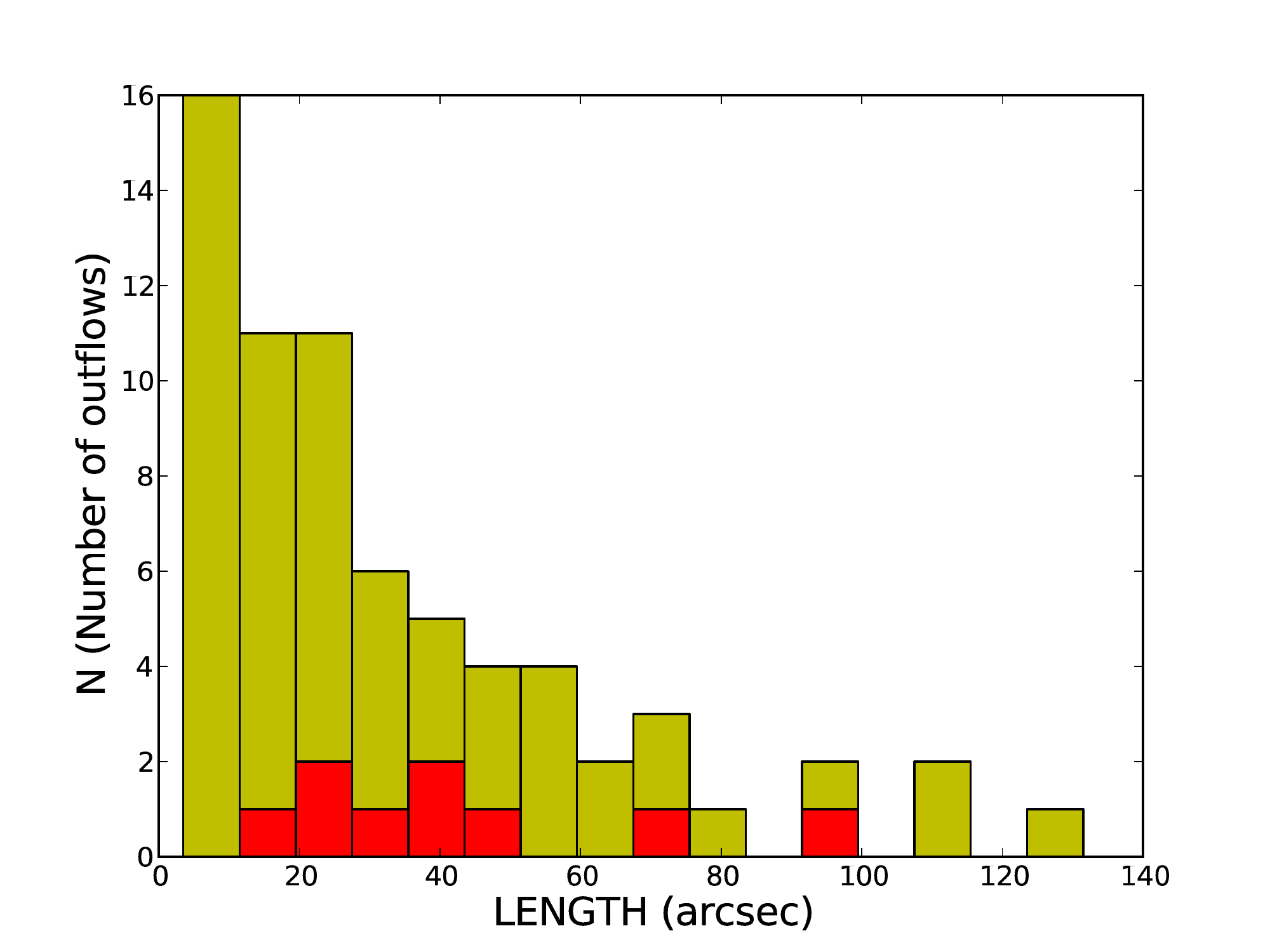}
\caption{\label{lengths} Histogram of the apparent outflow lengths. The light
grey areas show all outflows, while the dark grey areas represent the known
objects. In case there are several source candidates for one flow, the length
corresponding to the most likely source is plotted. Note that an apparent
length of 60\arcsec corresponds to about 0.9\,pc for the assumed distance of
3\,kpc.}
\end{figure}

For all outflows with assigned source candidate(s) we have measured the apparent
length(s). A histogram of the resulting lengths distribution is shown in
Fig.\,\ref{lengths}. The light grey bars represent all the outflows (68), while
the dark grey areas represent the already known objects (9). Note that the
lengths are not corrected for the unknown inclination angles or the potential 
detection of just parts of the outflow. For single sided flows the lengths 
correspond to the separation of the source and the most distant H$_2$ feature, 
while for bipolar flows the lengths correspond to the total length of the 
flow, a procedure also adopted in other surveys \citep{Stanke2002}. The known flows
do not occupy a special part of the diagram, but rather represent a random
sub-sample. Note that in cases of several source candidates for an outflow, we
plot the distance corresponding to the most likely source. The distribution will
not change significantly if we plot any other lengths.

There is a large range of apparent outflow lengths (from 4\arcsec\ to
130\arcsec) and a strong decrease of the number of outflows with increasing
length. Of the 68 outflows, more than 60\,\% have an apparent length of less
than 30\arcsec\ (or less than 0.4\,pc if we assume a distance of 3\,kpc). This
is in agreement with the distribution of lengths found by \cite{Davis2009} and
\cite{Stanke2002} along the Orion\,A molecular ridge and \cite{Davis2008} in
Taurus-Auriga-Perseus (NGC\,1333, L\,1455, L\,1448 and B\,1). Only 10\,\% of 
all outflows have a length of more than 1\,pc (if they are at 3\,kpc). In
Paper\,II we will discuss the physical outflow length distribution in parsec
after considering the individual distances to the objects. Note that the
 observed distribution does not agree with a randomly 
orientated sample of flows which all have the same length.

\subsection{Outflow position angles}

Previous studies (e.g. \cite{Mouschovias1976}) suggested that clouds collapse 
along magnetic field lines to form elongated, clumpy filaments from which 
chains of protostars are born. Furthermore, the associated outflows are 
aligned parallel with the local magnetic field and perpendicular to the chains 
of cores \citep{Banerjee2006}. Our sample is ideal to study the relation 
between cores, filaments and outflows because of the large sample of flows 
that are distributed over a considerable number of molecular clouds. We hence 
determined the outflow position angles and list them in 
Table\,\ref{outflow_table} in the Appendix.

\begin{figure}
\includegraphics[width=8.5cm]{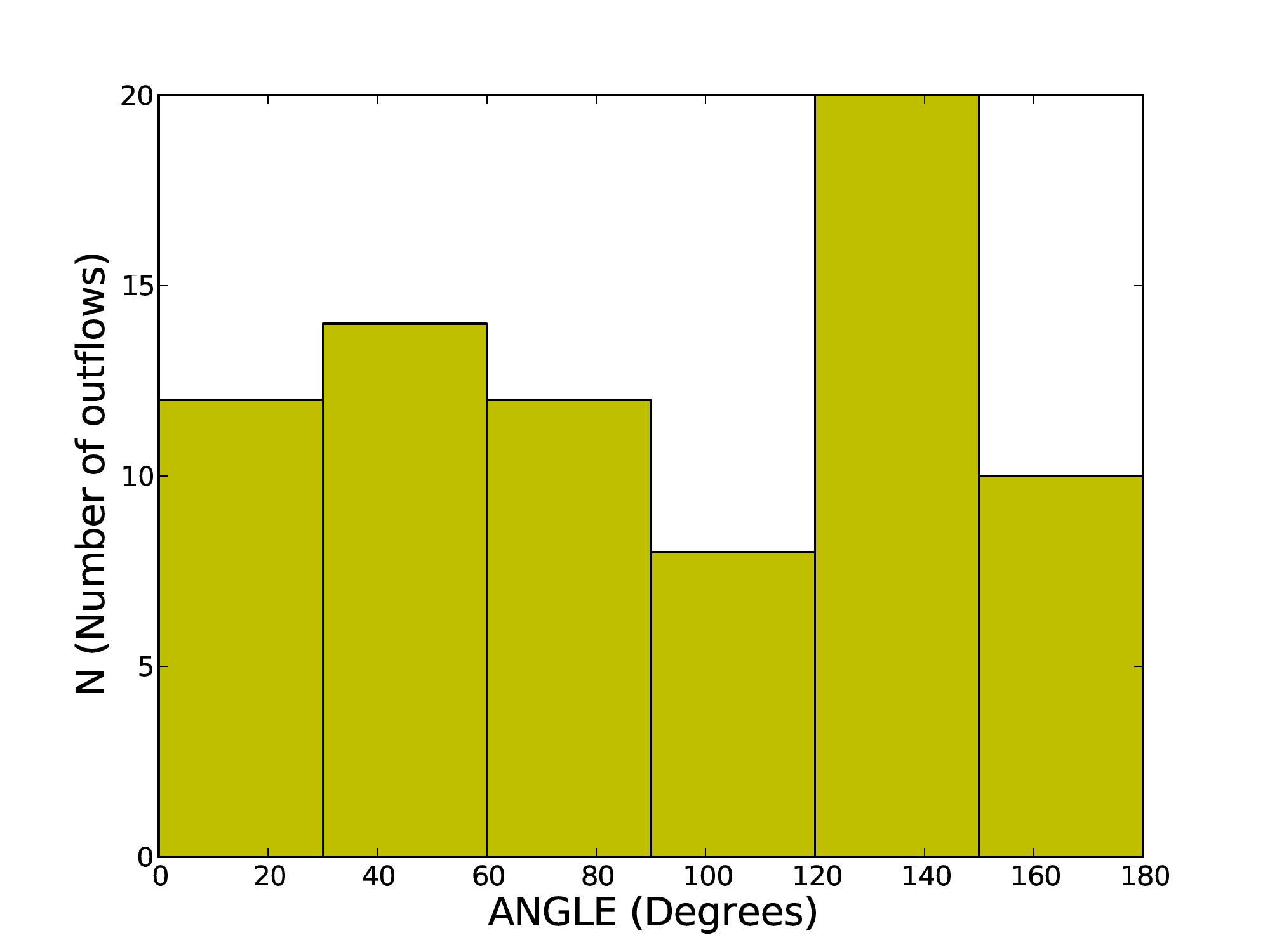}
\caption{\label{angle} Histogram of the distribution of position angles from
the detected outflows in 30 degrees bins. The distribution shows a 2.6\,$sigma$ over abundance
of flows with position angles between $l$\,=\,120$^\circ$ and
$l$\,=\,150$^\circ$, or an orientation roughly perpendicular to the Galactic Plane.}
\end{figure}

A histogram of the position angles is shown in Fig.\,\ref{angle}, using a 
bin-size of 30 degrees. The plot shows that a large fraction of outflows has a 
homogeneous distribution of position angles, with five of the six bins 
occupied by 11.2 objects on average. However, at angles between 120$^\circ$ 
and 150$^\circ$ one can identify an over abundance of outflows. There are 20 
outflows in this bin, which corresponds to a 2.6$\sigma$ deviation from the 
average of the other bins. We performed a Kolmogorov-Smirnov test to determine 
the probability that the observed position angle distribution is identical to 
a homogeneously distributed sample. For outflows with two potential sources, 
and hence two possible position angles, each angle was weighted by half; if 
there are three sources the weight for each angle was one third. We find a 
probability of just 10\,\% that our objects are drawn from a homogeneously 
distributed sample.

We further investigated if there is a trend of objects with a particular range 
of position angles with sky position. Nothing could be found. The objects 
falling into a particular bin in the position angle histogram are distributed 
completely homogeneously amongst the entire sample.

The Galactic plane has an inclination of about 64 degrees with respect to the 
ecliptic in our survey area. Thus, the peak in the position angle histogram 
between 120$^\circ$ and 150$^\circ$ indicates that there is a 2.6$\sigma$ over 
abundance of outflows orientated almost (PA\,=\,20$^\circ$ with respect to 
Galactic North) perpendicular to the Galactic Plane. It is therefore possible 
that our measured flow position angles are influenced by the large scale 
cloud structure. Previous studies of the Orion\,A molecular cloud by 
\cite{Davis2009} and \cite{Stanke2002} and in Taurus-Auriga-Perseus by 
\cite{Davis2008} have shown a homogeneous distribution with no significant 
trends in the orientation of outflows. However, studies of DR\,21/W\,75 by 
\cite{Davis2007} have shown that flows (in particular massive ones) are 
orthogonal to some degree to the molecular ridge. In Paper\,III we will 
analyse this distribution in more detail. We will measure, if possible, the 
orientation of the outflows with respect to its parental cloud filament, to 
verify or disprove the above findings.

\subsection{Outflow flux distribution}\label{resflux}
 
The measured integrated 1-0\,S(1) fluxes of the outflows are listed in
Table\,\ref{outflow_table} in the Appendix. Figure\,\ref{flux} shows a histogram
of this flux distribution. The light grey areas include all the outflows while
the dark grey areas are the already known objects. Clearly, the known outflows
dominate the bright end of the distribution and all but one are brighter than
10$^{-17}$\,W\,m$^{-2}$. In particular, about 60\,\% of the integrated
flux in the 1-0\,S(1) line of molecular hydrogen can be attributed to already
known MHOs. Thus, our survey just about doubles the known integrated
1-0\,S(1) flux from jets and outflows in the survey area, in contrast to the 15
fold increase in total outflow numbers.

In general we find a sharp rise of the number of MHOs with decreasing
integrated flux. More than 70\,\% of the MHOs have an integrated flux of less
than 10$^{-17}$\,W\,m$^{-2}$. The maximum number of MHOs occurs at about
3\,$\cdot$\,10$^{-18}$\,W\,m$^{-2}$ which corresponds to about 10$^{-3}$ solar
luminosities in the 1-0\,S(1) line of H$_2$ at our assumed distance of 3\,kpc.
We hence conclude that this flux is the completeness limit for outflows detected
in our survey. There are, however, a number of objects with a significantly
smaller integrated flux, indicating that the actual detection limit for
1-0\,S(1) fluxes is lower than even 10$^{-18}$\,W\,m$^{-2}$ in some regions
(most likely dark clouds devoid of fore/background stars). This is also
supported by fact that many of the outflows with a small overall integrated
flux consist of only one emission line feature, 

For each of the integrated fluxes we list the photometric uncertainties in
Table\,\ref{outflow_table}. Typical uncertainties of the measured fluxes are of
the order of 10\,\%. For bright objects they are significantly smaller. Wherever
the integrated flux is very low and the variation of the local background
level is high we only determine an upper flux limit.

How the measured integrated flux distribution converts into a statistically
corrected luminosity distribution will be discussed in a forthcoming paper (Paper\,II).

\begin{figure}
\includegraphics[width=8.5cm]{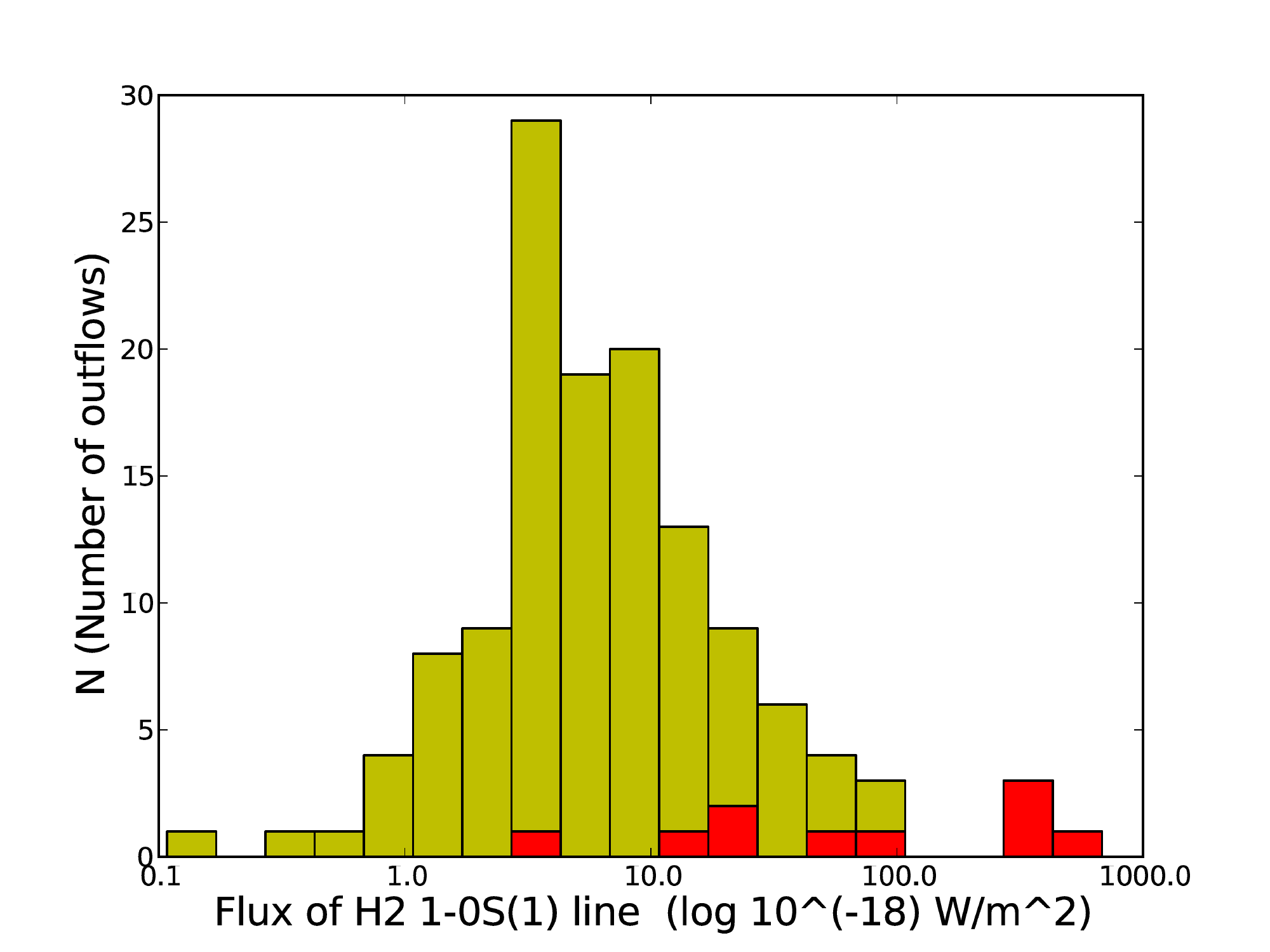}
\caption{\label{flux} Histogram of integrated 1-0\,S(1) fluxes (in logarithmic
units) for the detected outflows. The light grey areas include all outflows
while the dark grey areas represent the already known MHOs. Our completeness
limit is about 3\,$\cdot$\,10$^{-18}$\,W\,m$^{-2}$. }
\end{figure}

\section{Conclusions}\label{conclusions}

In order to investigate the dynamical component of the star formation process we
perform an unbiased search for jets and outflows from YSOs along the Galactic
Plane. Our data has been taken as part of the UWISH2 survey (F11). It uses as a
tracer the 1-0\,S(1) emission line of H$_2$, and here we focus our attention on
a continuous 33 square degree sized region (18\dg \textless l \textless 30\dg;
-1.5\dg \textless b \textless +1.5\dg) in Serpens and Aquila.

We identify 131 outflows from YSOs from which 94\,\% (123 objects) are new
discoveries in our data set. Therefore, our survey has increased the number of
known molecular hydrogen outflows by a factor of 15 in the area investigated.

We find a flux completeness limit for our outflow detection of
3\,$\cdot$\,10$^{-18}$\,W\,m$^{-2}$, with 70\,\% of the objects showing fluxes
of 10$^{-17}$\,W\,m$^{-2}$ or less. Typically, the already known outflows occupy
the bright end of the flux distribution. Our survey thus increases the  
known integrated 1-0\,S(1) H$_2$ flux from jets and outflows only by a factor
of two, compared to the large increase in the total number of flows. 

The overall spatial distribution of the detected outflows shows that they follow
the GMC complexes in the Galactic Plane. Only a small number of objects is not
situated within dense clouds. However, there are large areas with high
extinction where no outflows have been detected. Overall, about 2/3rd of the
survey area is more or less completely devoid of jets and outflows. We further
find that the flows typically occur in groups of 3\,--\,5 members with a size of
about 5\,pc (at our assumed distance of 3\,kpc). These groups are typically
separated by about half a degree on the sky. 

The distribution of flows perpendicular to the Galactic Plane shows a
Gaussian like distribution with a width of about 25\,--\,30\,pc (at our assumed
distance of 3\,kpc), similar to values for young stellar clusters and OB stars.

We are able to assign possible driving sources to about 50\,\% of the outflows.
Brighter MHOs are more likely to have a source candidate assigned to them. 

We measure the apparent outflow length for outflows with assigned driving
sources. There is a wide range of lengths from 4\arcsec\ to 130\arcsec\ and a
strong decrease of the number of flows with increasing length. More than 60\,\%
of the outflows have an apparent length of less than 30\arcsec\  (or less than
0.4\,pc if we assume a distance of 3\,kpc) while parsec-scale outflows are not
common. Only 10\,\% of all outflows would have a length of greater than 1\,pc at our
assumed distance.

The position angle distribution of flows with assigned source shows an
2.6\,$\sigma$ over-abundance at angles between 120$^\circ$ and 150$^\circ$.

\section*{acknowledgements}

GI acknowledges a University of Kent scholarship. The United Kingdom Infrared
Telescope is operated by the Joint Astronomy Centre on behalf of the Science and
Technology Facilities Council of the U.K. The data reported here were obtained
as part of the UKIRT Service Program. This research has made use of the WEBDA
database, operated at the Institute for Astronomy of the University of Vienna.

\bibliographystyle{mn2e}
\bibliography{serpens1_complete}

\newpage
\onecolumn

\begin{appendix}

\begin{landscape}

\section{MHO Table}\label{appendix}

\begin{center}

\begin{longtable}{|l|l|l|c|c|c|c|l|l|l|}

\caption{Summary table of the MHO properties. The table lists the MHO number, 
Right Ascension and Declination (J2000), the integrated 1-0\,S(1) flux 
and its uncertainty, the apparent length, the position angle, the source 
candidate(s) and its position(s). In cases where several MHOs belong to the 
same outflow, the MHO number is labelled with an asterisk. If the flux is an 
upper limit, no flux errors are given. Only for objects with a source 
candidate, we determined the apparent length and the position angle. Flow 
lengths are end-to-end for bipolar outflows or source-to-end for single sided 
flows. Source names correspond to the catalogue entries the source was taken 
from. These are either coordinate based, SIMBAD names, UKIDSS GPS IDs (12 
digit numbers), or AKARI IDs (7 digit numbers). Sources labelled 'Noname' are 
objects where it is assumed that there is a driving source at the listed 
position, however, non of the catalogues used shows a detection of the object. 
Please see the main text for details on the values in the table.}

\label{outflow_table}\\

\hline \multicolumn{1}{|c|}{\textbf{MHO}} & \multicolumn{1}{c|}{\textbf{RA}} & \multicolumn{1}{c|}{\textbf{DEC}} & \multicolumn{1}{c|}{\textbf{F[1-0\,S(1)]}} &
\multicolumn{1}{c|}{\textbf{Flux error}}
& \multicolumn{1}{c|}{\textbf{length}} & \multicolumn{1}{c|}{\textbf{position angle}} & \multicolumn{1}{c|}{\textbf{possible}} & \multicolumn{1}{c|}{\textbf{source RA}} & \multicolumn{1}{c|}{\textbf{source DEC}}\\
\multicolumn{1}{|c|}{\textbf{ }} & \multicolumn{1}{c|}{\textbf{(J2000)}} & \multicolumn{1}{c|}{\textbf{(J2000)}} & \multicolumn{1}{c|}{\textbf{[10E-18 W/m$^2$]}} & \multicolumn{1}{c|}{\textbf{[10E-18 W/m$^2$]}}
& \multicolumn{1}{c|}{\textbf{(arcsec)}} & \multicolumn{1}{c|}{\textbf{(degrees)}} & \multicolumn{1}{c|}{\textbf{source}} & \multicolumn{1}{c|}{\textbf{(J2000)}} & \multicolumn{1}{c|}{\textbf{(J2000)}}\\ \hline 
\endfirsthead

\multicolumn{10}{c}%
{{\bfseries \tablename\ \thetable{} -- continued from previous page}} \\
\hline \multicolumn{1}{|c|}{\textbf{MHO}} & \multicolumn{1}{c|}{\textbf{RA}} & \multicolumn{1}{c|}{\textbf{DEC}} & \multicolumn{1}{c|}{\textbf{F[1-0\,S(1)]}} &
\multicolumn{1}{c|}{\textbf{Flux error}}
& \multicolumn{1}{c|}{\textbf{length}} & \multicolumn{1}{c|}{\textbf{position angle}} & \multicolumn{1}{c|}{\textbf{possible}} & \multicolumn{1}{c|}{\textbf{source RA}} & \multicolumn{1}{c|}{\textbf{source DEC}}\\
\multicolumn{1}{|c|}{\textbf{ }} & \multicolumn{1}{c|}{\textbf{(J2000)}} & \multicolumn{1}{c|}{\textbf{(J2000)}} & \multicolumn{1}{c|}{\textbf{[10E-18 W/m$^2$]}} & \multicolumn{1}{c|}{\textbf{[10E-18 W/m$^2$]}}
& \multicolumn{1}{c|}{\textbf{(arcsec)}} & \multicolumn{1}{c|}{\textbf{(degrees)}} & \multicolumn{1}{c|}{\textbf{source}} & \multicolumn{1}{c|}{\textbf{(J2000)}} & \multicolumn{1}{c|}{\textbf{(J2000)}}\\ \hline 
\endhead

\hline \multicolumn{10}{|r|}{{Continued on next page}} \\ \hline
\endfoot

\hline \hline
\endlastfoot

MHO 2201*& 18:17:57.7 & -12:07:19 & 379.119 & 37.885  & 73  & 132 & IRAS 18151-1208  & 18:17:57.9 & -12:07:20\\
MHO 2202 & 18:17:57.2 & -12:07:30 & 77.481  & 5.632   & 21  & 32  & IRAS 18151-1208  & 18:17:57.9 & -12:07:20\\
MHO 2203 & 18:29:16.6 & -11:50:17 & 278.893 & 19.586  & 42 & 67   & G019.884-00.535   & 18:29:14.7 & -11:50:24\\
MHO 2204 & 18:29:13.0 & -11:50:16 & 509.659 & 42.892  & 50 & 116  & G019.884-00.535   & 18:29:14.7 & -11:50:24\\
MHO 2205 & 18:29:16.9 & -11:49:54 & 58.429  & 3.918   & -  & 29   & -		    & - 	 & -	    \\
MHO 2206*& 18:34:22.7 & -05:59:59 & 286.000 & 18.607  & 93  & 131 & IRAS 18316-0602  & 18:34:20.9 & -05:59:42\\
MHO 2207*& 18:34:20.5 & -05:59:38 & 9.664   & 1.027   & *   & *   & *		   & *  	& *	   \\
MHO 2208*& 18:34:18.4 & -05:59:07 & 8.720   & 0.609   & *   & *   & *		   & *  	& *	   \\
MHO 2209 & 18:34:18.8 & -05:59:25 & 25.340  & 1.982   & 26  & 137 & 438649130203	    & 18:34:20.0 & -05:59:46\\
MHO 2210 & 18:34:21.3 & -06:00:12 & 20.150  & 1.260   & 40  & 168 & IRAS 18316-0602  & 18:34:20.9 & -05:59:42\\
MHO 2212*& 18:17:55.4 & -12:06:42 & 26.363  & 1.926   & -   & -   & -		    & - 	 &	   -\\
MHO 2244 & 18:25:44.8 & -12:22:46 & 3.005   & 0.217   & 35 & 160  & 3329252	   & 18:25:44.7 & -12:22:34\\
MHO 2245 & 18:29:14.8 & -11:50:08 & 12.587  & 0.733   & 18 & 7	  & G019.884-00.535   & 18:29:14.7 & -11:50:24\\
MHO 2246 & 18:29:12.7 & -11:50:31 &  7.588  & 1.385   & 10 & 55	  & G019.8810-00.5300 & 18:29:13.3 & -11:50:25\\
MHO 2247 & 18:26:58.9 & -11:31:47 & 19.772  & 9.532   & 70 & 45   & G019.896+00.103   & 18:26:57.6 & -11:31:58\\
MHO 2248 & 18:29:12.3 & -11:47:53 & 3.800   & 2.065   & 26 & 88   & 438521940668      & 18:29:13.2 & -11:47:51\\
MHO 2249 & 18:30:04.6 & -11:55:34 & 3.669   & 0.213   & 95 & 95   & G019.9122-00.7799 & 18:30:11.1 & -11:55:42\\
         &            &           &         & 	      & 64 & 145  & 3318624	    & 18:30:06.9 & -11:56:28\\
MHO 2250 & 18:28:18.9 & -11:39:28 & 8.288   & 0.539   & 29 & 133  & G019.9357-00.2558 & 18:28:19.9 & -11:39:52\\
MHO 2251 & 18:28:19.1 & -11:40:39 & 3.033   & 0.427   & 4  & 130  & 438521818214     & 18:28:18.9 & -11:40:36\\
MHO 2252 & 18:25:22.8 & -12:54:52 & 2.406   & 0.229   & 8.5& 19   & 3349808	   & 18:25:22.7 & -12:55:00\\
MHO 2253 & 18:25:54.7 & -12:01:59 & 10.497  & 0.640   & -  & -	  & -		   & -  	& -	   \\
MHO 2254 & 18:27:12.6 & -10:33:57 & 9.463   & 0.543   & 54 & 44   & 3054756	   & 18:27:15.2 & -10:33:20\\
MHO 2255 & 18:29:10.9 & -10:18:07 & 1.058   & 0.279   & 10 & 66   & G021.2369+00.1940 & 18:29:10.2 & -10:18:11\\
MHO 2256 & 18:29:09.3 & -10:18:24 & 3.435   & 0.277   & 9  & 68   & Noname	   & 18:29:09.3 & -10:18:24\\
MHO 2257 & 18:29:08.5 & -10:18:54 & 4.862   & 0.751   & 18 & 9	  & G021.2248+00.1953 & 18:29:08.6 & -10:18:48\\
MHO 2258 & 18:32:17.1 & -10:34:19 & 4.044   & 0.428   & -  & -	  & -		   & -  	& -	   \\
MHO 2259 & 18:29:03.7 & -09:44:13 & 7.424   & 0.867   & -  & -	  & -		   & -  	& -	   \\
MHO 2260 & 18:30:35.0 & -09:34:45 & 22.339  & 1.768   & 25 & 78   & 438379615372	   & 18:30:33.5 & -09:34:49\\
MHO 2261 & 18:30:33.4 & -09:35:07 & 62.588  & 7.425   & 72 & 29   & 438379615655	   & 18:30:34.0 & -09:34:52\\
MHO 2262 & 18:33:20.1 & -07:55:44 & 15.495  & Upper Limit  & 18   & 142 & 3058500	   & 18:33:19.4 & -07:55:32\\
MHO 2263 & 18:34:20.7 & -08:17:51 & 6.439   & 1.059   & 23 & 166  & G023.6008-00.0147 & 18:34:21.1 & -08:18:13\\
MHO 2264 & 18:34:23.4 & -08:18:07 & 15.185  & 1.073   & 18 & 40   & 3057470	   & 18:34:22.7 & -08:18:20\\
MHO 2265 & 18:33:11.6 & -10:06:15 & 3.983   & 0.469   & 18 & 100  & 438517255051     & 18:33:12.7	& -10:06:18 \\
         &            &           &         &         & 5  & 136  & 438517254983	   & 18:33:11.4 & -10:06:12\\
MHO 2266 & 18:30:41.7 & -09:34:58 & 5.976   & 0.509   & 28 & 0	  & G022.0535+00.1980 & 18:30:41.6 & -09:34:40\\
MHO 2267 & 18:34:51.3 & -08:57:51 & 3.236   & 0.246   & -  & -	  & -		   & -  	& -	   \\
MHO 2268 & 18:34:21.3 & -08:33:24 & 0.111   & Upper Limit  & -    & -	& -		   & -  	& -	   \\
MHO 2269 & 18:34:31.0 & -08:42:48 & 32.477  & 5.279   & 60 & 77   & JCMTSE J183431.5-084250 & 18:34:31.4 & -08:42:46\\
MHO 2270 & 18:35:28.5 & -08:51:43 & 1.148   & 0.099   & 8  & 158  & Noname	   & 18:35:28.3 & -08:51:36\\
MHO 2271 & 18:35:31.7 & -08:52:17 & 6.578   & 0.465   & -  & -	  & -		   & -  	& -	   \\
MHO 2272 & 18:35:51.3 & -08:41:13 & 3.822   & 0.239   & 4  & 45   & G023.4319-00.5212 & 18:35:51.4 & -08:41:10\\
MHO 2273 & 18:35:07.0 & -08:05:47 & 2.535   & 0.338   & -  & -	  & -		   & -  	& -	   \\
MHO 2274 & 18:36:41.1 & -07:39:20 & 17.829  & 2.088   & 66 & 0	  & 438511042387	   & 18:36:40.9 & -07:39:02\\
MHO 2275 & 18:36:13.0 & -07:42:40 & 17.181  & 2.264   & -  & -	  & -		   & -  	& -	   \\
MHO 2276 & 18:35:22.9 & -07:19:17 & 2.777   & 0.165   & -  & -	  & -		   & -  	& -	   \\
MHO 2277 & 18:33:51.0 & -07:06:13 & 24.144  & 1.847   & -  & 134  & -		   & -  	& -	   \\
MHO 2278 & 18:36:09.2 & -06:50:07 & 13.823  & 0.936   & 20 & 36   & 3338999	   & 18:36:09.9 & -06:49:53\\
MHO 2279 & 18:38:51.9 & -06:51:01 & 3.023   & 0.215   & 5  & 117  & 438745913239	   & 18:38:52.1 & -06:51:02\\
MHO 2280 & 18:38:55.4 & -06:52:37 & 6.770   & 1.073   & -  & -	  & -		   & -  	& -	   \\
MHO 2281 & 18:38:57.3 & -06:53:15 & 6.350   & 0.620   & -  & -	  & -		   & -  	& -	   \\
MHO 2282 & 18:36:56.6 & -06:48:24 & 3.561   & 0.233   & -  & -	  & -		   & -  	& -	   \\
MHO 2283 & 18:37:21.9 & -07:31:56 & 2.511   & 0.242   & 20 & 39   & 438368016438     & 18:37:22.7 & -07:31:41\\
         &            &           &         &         & 34 & 103  & G024.6280-00.3175 & 18:37:20.7 & -07:31:51\\
MHO 2284 & 18:37:22.9 & -07:31:58 & 3.745   & 0.674   & 18 & 170  & 438368016438	      & 18:37:22.7 & -07:31:41\\
         &            &           &         &         & 20 & 12   & G024.635-00.325   & 18:37:23.3 & -07:31:40\\
         &            &           &         &         & 34 & 103  & G024.6280-00.3175 & 18:37:20.7 & -07:31:51\\
MHO 2285 & 18:37:22.1 & -07:32:14 & 2.482   & 0.183   & 33 & 14   & 438368016438      & 18:37:22.7 & -07:31:41\\
         &            &           &         &         & 33 & 139  & G024.6280-00.3175 & 18:37:20.7 & -07:31:51\\
         &            &           &         &         & 38 & 25   & G024.635-00.325   & 18:37:23.3 & -07:31:40\\
MHO 2286 & 18:37:22.4 & -07:37:08 & 13.093  & 1.088   & -  & 126  & -		    & - 	 & -	    \\
MHO 2287 & 18:37:12.3 & -07:11:26 & 3.239   & 0.192   & -  & -	  & -		    & - 	 & -	    \\
MHO 2288 & 18:37:21.6 & -07:11:38 & 10.087  & 0.622   & -  & -	  & -		    & - 	 & -	    \\
MHO 2289 & 18:37:21.7 & -07:11:46 & 3.036   & 0.182   & -  & -	  & -		    & - 	 & -	    \\
MHO 2290 & 18:37:19.8 & -07:11:44 & 12.735  & 1.197   & 27 & 140  & G024.926-00.157   & 18:37:19.4 & -07:11:35\\
MHO 2291 & 18:39:11.3 & -07:20:09 & 38.832  & 3.432   & 112 & 147 & 438366899733	    & 18:39:11.8 & -07:20:22\\
MHO 2292 & 18:38:59.6 & -06:12:44 & 7.950   & 0.622   & 6   & 147 & Noname	    & 18:38:59.6 & -06:12:44\\
MHO 2293 & 18:37:15.9 & -05:19:25 & 5.838   & 0.581   & -   & -   & -		    & - 	 & -	    \\
MHO 2294 & 18:37:05.4 & -05:24:02 & 1.553   & 0.121   & -   & -   & -		    & - 	 & -	    \\
MHO 2295 & 18:37:04.4 & -05:24:02 & 1.385   & 0.284   & -   & -   & -		    & - 	 & -	    \\
MHO 2296 & 18:37:08.7 & -05:24:08 & 2.170   & 0.125   & -   & -   & -		    & - 	 & -	    \\
MHO 2297 & 18:38:59.8 & -05:27:06 & 1.326   & 0.320   & 10  & 114 & 438361450423	    & 18:38:59.2 & -05:27:03\\
MHO 2298 & 18:39:50.0 & -04:30:46 & 30.286  & 3.048   & -   & -   & -		    & - 	 & -	    \\
MHO 2299 & 18:25:54.5 & -10:12:34 & 4.781   & 0.358   & -   & -   & -		    & - 	 & -	    \\
MHO 2436 & 18:40:21.9 & -04:27:56 & 10.940  & 1.717   & -   & -   & -		    & - 	 & -	    \\
MHO 2437 & 18:41:12.1 & -04:56:43 & 28.685  & 5.722   & -   & -   & -		    & - 	 & -	    \\
MHO 2438 & 18:41:51.9 & -04:19:53 & 5.005   & 0.383   & -   & -   & -		    & - 	 & -	    \\
MHO 2439 & 18:44:08.4 & -04:33:27 & 5.294   & 0.316   & -   & 51  & -		    & - 	 & -	    \\
MHO 2440 & 18:44:08.6 & -04:33:27 & 6.636   & 0.513   & 20  & 160 & G028.0473-00.4562 & 18:44:08.4 & -04:33:17\\
MHO 2441 & 18:44:03.6 & -04:38:03 & 19.608  & 1.847   & 12  & 73  & G027.97-0.47	    & 18:44:03.7 & -04:38:02\\
MHO 2442 & 18:43:11.0 & -04:41:15 & 13.718  & 1.164   & -   & -	  & -		    & - 	 & -	    \\
MHO 2443 & 18:44:44.4 & -04:13:30 & 7.008   & 0.463   & -   & -	  & -		    & - 	 & -	    \\
MHO 2444 & 18:42:58.4 & -04:08:04 & 11.168  & 0.710   & 3.5 & 152 & J184259.0-040802  & 18:42:58.4 & -04:08:04 \\
MHO 2445 & 18:43:08.6 & -04:18:32 & 3.482   & 0.203   & 11  & 143 & 438646853052	    & 18:43:09.0 & -04:18:40\\
MHO 2446 & 18:41:33.6 & -04:01:22 & 3.608   & 0.432   & -   & -	  & -		    & - 	 & -	    \\
MHO 2447 & 18:41:33.6 & -04:01:41 & 6.586   & 1.695   & -   & -	  & -		    & - 	 & -	    \\
MHO 2448 & 18:42:51.7 & -03:59:35 & 7.518   & 0.544   & 115 & 135 & 3344177	    & 18:42:52.5 & -03:59:47\\
MHO 2449 & 18:42:56.3 & -03:59:52 & 3.770   & 0.204   & 4   & 157 & Noname           & 18:42:56.3 & -03:59:51\\
MHO 2450 & 18:42:49.7 & -04:02:22 & 0.680   & Upper Limit   & 15  & 20  & G028.3566+00.0694& 18:42:49.9 & -04:02:22\\
MHO 2451 & 18:42:54.0 & -04:02:28 & 8.806   & 0.572   & -   & 35  & -		    & - 	 & -	    \\
MHO 2452 & 18:42:49.9 & -04:02:46 & 0.719   & 0.102   & -   & -   & -		    & - 	 & -	    \\
MHO 2453 & 18:42:52.3 & -04:03:21 & 3.087   & 0.291   & 43  & 108 & G028.3450+00.0628& 18:42:50.0 & -04:03:09\\
MHO 2454 & 18:43:01.5 & -03:34:49 & 45.166  & 4.229   & 25  & 146 & 438742594697	    & 18:43:01.6 & -03:34:51\\
MHO 2455 & 18:43:18.6 & -03:35:25 & 5.353   & 0.642   & -   & -   & -		    & - 	 & -	    \\
MHO 2456 & 18:41:11.2 & -07:51:31 & 8.493   & 0.763   & -   & -   & -		    & - 	 &	   -\\
MHO 3200 & 18:26:20.2 & -10:13:42 & 68.260  & 6.839   & 80  & 134 & G20.9766+00.8367 & 18:26:21.9 & -10:14:05\\
MHO 3201 & 18:26:21.6 & -10:14:07 & 3.534   & 0.949   & 7   & 70  & G20.9766+00.8367 & 18:26:21.9 & -10:14:05\\
MHO 3202 & 18:33:49.3 & -05:16:16 & 6.238   & 0.514   & 51  & 88  & Noname	    & 18:33:49.4 & -05:16:15\\
MHO 3203 & 18:33:30.6 & -05:00:35 & 3.383   & 0.328   & -   & -   & -		    & - 	 & -	    \\
MHO 3204 & 18:33:38.9 & -05:01:16 & 3.304   & 0.229   & 50  & 46  & 438744307809	    & 18:33:37.3 & -05:01:37\\
MHO 3205 & 18:33:34.2 & -05:10:25 & 2.275   & 0.171   & 12  & 95  & 438744300302	    & 18:33:33.8 & -05:10:24\\
MHO 3206 & 18:34:08.7 & -05:15:00 & 3.439   & 0.221   & -   & -   & -		    & - 	 & -	    \\
MHO 3207 & 18:34:12.1 & -05:15:14 & 4.159   & 0.269   & -   & -   & -		    & - 	 & -	    \\
MHO 3208 & 18:34:03.9 & -05:16:18 & 1.440   & 0.100   & -   & -   & -		    & - 	 & -	    \\
MHO 3209 & 18:34:10.3 & -05:16:52 & 5.022   & 0.582   & -   & -   & -		    & - 	 & -	    \\
MHO 3210 & 18:34:17.9 & -05:04:34 & 2.853   & 0.204   & 12  & 47  & 438744216219	    & 18:34:18.2 & -05:04:28\\
MHO 3211 & 18:34:04.1 & -05:06:01 & 29.166  & 2.093   & 54  & 63  & Noname	    & 18:34:04.0 & -05:06:01\\
MHO 3212 & 18:34:13.0 & -05:07:33 & 2.527   & 0.552   & -   & -   & -		    & - 	 & -        \\
MHO 3213 & 18:34:42.1 & -05:08:55 & 6.857   & 0.887   & 34  & 14  & Noname	    & 18:34:42.1 & -05:08:55\\
MHO 3214 & 18:34:24.9 & -05:10:58 & 6.214   & 2.819   & 28  & 82  & Noname	    & 18:34:25.6 & -05:10:57\\
MHO 3215 & 18:34:30.0 & -05:11:10 & 4.387   & 0.607   & -   & -   & -	            & -          & -        \\
MHO 3216 & 18:17:50.2 & -12:07:55 & 82.356  & 6.942   & 26  & 128 & 438753255942	    & 18:17:50.9 & -12:08:03\\
MHO 3217 & 18:31:09.4 & -12:30:03 & 15.238  & 1.667   & 45  & 31  & 438386549587	    & 18:31:09.3 & -12:30:05\\
MHO 3218 & 18:30:51.5 & -12:33:05 & 5.549   & 0.492   & 58  & 147 & 438386550198	    & 18:30:53.2 & -12:33:43\\
MHO 3219 & 18:30:53.7 & -12:33:31 & 12.049  & 0.894   & 59  & 44  & 438386550198	    & 18:30:53.2 & -12:33:43\\
MHO 3220 & 18:30:54.5 & -12:33:44 & 8.344   & 1.212   & -   & 159 & -		    & - 	 & -	    \\
MHO 3221 & 18:29:37.1 & -12:37:18 & 16.329  & 1.314   & -   & -   & -		    & - 	 &-	    \\
MHO 3222 & 18:30:30.6 & -11:44:28 & 10.380  & 0.659   & 6.5 & 149 & Noname	    & 18:30:30.6 & -11:44:28\\
MHO 3223 & 18:30:44.8 & -11:49:35 & 3.129   & 0.268   & -   & -   & -		    & - 	 &	   -\\
MHO 3224 & 18:31:49.7 & -11:52:51 & 1.700   & 0.178   & -   & -   & -		    & - 	 &	   -\\
MHO 3225 & 18:31:41.5 & -11:53:54 & 0.964   & 0.051   & -   & -   & -		    & - 	 &	   -\\
MHO 3226 & 18:31:49.5 & -11:54:12 & 3.466   & 0.360   & -   & -   & -		    & - 	 &	   -\\
MHO 3227 & 18:31:41.4 & -11:55:13 & 0.586   & 0.031   & -   & -   & -		    & - 	 &	   -\\
MHO 3228 & 18:19:56.4 & -11:33:30 & 2.092   & 0.143   & -   & -   & -		    & - 	 &	   -\\
MHO 3229 & 18:38:03.0 & -07:44:19 & 3.135   & 0.649   & -   & 70  & -		    & - 	 &	   -\\
MHO 3230 & 18:20:02.2 & -11:34:23 & 5.628   & 0.395   & -   & -   & -		    & - 	 &	   -\\
MHO 3231 & 18:19:56.9 & -11:34:25 & 2.024   & 0.232   & -   & -   & -		    & - 	 &	   -\\
MHO 3232 & 18:19:55.3 & -11:34:45 & 1.201   & 0.074   & -   & -   & -		    & - 	 &	   -\\
MHO 3233 & 18:20:09.8 & -11:18:52 & 1.428   & 0.080   & -   & -   & -		    & - 	 &	   -\\
MHO 3234 & 18:34:51.9 & -10:44:34 & 3.940   & 0.321   & -   & -   & -		    & - 	 &	   -\\
MHO 3235 & 18:36:25.9 & -08:38:33 & 8.045   & 0.717   & -   & -   & -		    & - 	 &	   -\\
MHO 3236 & 18:29:36.8 & -07:43:59 & 0.913   & 0.074   & -   & -   & -		    & - 	 &	   -\\
MHO 3237 & 18:38:01.9 & -07:44:26 & 7.729   & 0.519   & 18  & 122 & G024.5201-00.5642& 18:38:01.8 & -07:44:24\\
MHO 3238 & 18:38:42.4 & -07:48:24 & 8.895   & 0.680   & -  & 150  & -		    & - 	 & -	    \\
MHO 3239 & 18:38:39.3 & -07:47:41 & 15.135  & 0.950   & -   & -   & -		    & - 	 &	   -\\
MHO 3240 & 18:38:27.2 & -07:49:11 & 60.245  & 5.703   & 126 & 13  & 438510698063	    & 18:38:27.1 & -07:49:12\\
MHO 3241 & 18:34:24.8 & -05:54:34 & 28.141  & 1.741   & 40  & 22  & Noname	    & 18:34:24.9 & -05:54:34\\
MHO 3242 & 18:34:17.2 & -06:00:24 & 16.002  & 0.912   & -   & -   & -		    & - 	 &	   -\\
MHO 3243 & 18:34:36.5 & -05:59:24 & 8.172   & 0.569   & 38  & 120 & 3060635	    & 18:34:38.7 & -05:59:38\\
MHO 3244 & 18:34:04.4 & -06:01:12 & 8.433   & 0.790   & -   & -   & -		    & - 	 &	   -\\
MHO 3246 & 18:31:08.3 & -10:20:43 & 17.575  & 1.737   & 4.5 & 102 & Noname           & 18:31:08.3 & -10:20:43\\

\end{longtable}

\end{center}

\end{landscape}

\newpage

\section{MHO images}

\begin{center}

\begin{longtable}{|r|c|p{3.5in}|}

\caption{This table contains close-up grey scale H$_2$-$K$ images for each
detected MHO in our paper, as well as a brief description of the morphology of
the emission. We further briefly mention sub-structure of the objects as well as
possible driving sources. All images are aligned with North to the top and East
to the left. A scale is indicated in every case. In the top right corner we
indicate the name of the H$_2$ image which contains the object. The red line
indicates the direction of the outflow while the red circles mark possible
driving sources. Labels on each driving source candidate refer to the
identification number in the source catalogue they are taken from (e.g. Glimpse,
Bolocam, IRAS, GPS; see text for details). MHOs that are marked with a * after
their MHO number are parts of one outflow.}

\label{images_table} \\

\hline \multicolumn{1}{|c|}{\textbf{MHO}} & \multicolumn{1}{c|}{\textbf{Image}} & \multicolumn{1}{c|}{\textbf{Comments}} \\ \hline 
\endfirsthead

\multicolumn{3}{c}%
{{\bfseries \tablename\ \thetable{} -- continued from previous page}} \\
\hline \multicolumn{1}{|c|}{\textbf{MHO}} & \multicolumn{1}{c|}{\textbf{Image}} & \multicolumn{1}{c|}{\textbf{Comments}} \\ \hline 
\endhead

\hline \multicolumn{3}{|r|}{{Continued on next page}} \\ \hline
\endfoot

\hline \hline
\endlastfoot

table not included & in & astro-ph version \\ \hline
\end{longtable}
\end{center}

\end{appendix}

\label{lastpage}

\end{document}